\documentclass{jpp}

\usepackage{graphicx}
\usepackage{natbib}
\usepackage{pdflscape}
\usepackage{bm}
\usepackage{float}

\def\Pm{\mbox{\rm P}_M}
\def\Rm{\mbox{\rm R}_M}
\def\Rmc{R_{\rm crit}}
\def\zr{\bar{z}}

\newcommand{\be}{\begin{equation}}
\newcommand{\ee}{\end{equation}}
\newcommand{\bea}{\begin{eqnarray}}
\newcommand{\eea}{\end{eqnarray}}

\newcommand{\Eq}[1]{Eq.~(\ref{#1})}
\newcommand{\Eqs}[2]{Eqs~(\ref{#1}) and~(\ref{#2})}
\newcommand{\Eeqs}[3]{Eqs~(\ref{#1}), (\ref{#2})~and~(\ref{#3})}

\newcommand{\bra}[1]{\langle #1\rangle}
\newcommand{\bbra}[1]{\left\langle #1\right\rangle}

\newcommand{\kk}{\mbox{\boldmath $k$}}
\newcommand{\pp}{\mbox{\boldmath $p$}}
\newcommand{\aaa}{\mbox{\boldmath $a$}}
\newcommand{\AAA}{\mbox{\boldmath $A$}}
\newcommand{\q}{\mbox{\boldmath $q$}}
\newcommand{\x}{\mbox{\boldmath $x$}}
\newcommand{\y}{\mbox{\boldmath $y$}}
\newcommand{\rr}{\mbox{\boldmath $r$}}
\newcommand{\rv}[1]{r_{#1}}
\newcommand{\rvu}[1]{\hat{r}_{#1}}
\newcommand{\PP}[1]{\hat{P}_{#1}}
\newcommand{\PPP}[1]{\tilde{P}_{#1}}
\newcommand{\uu}{\mbox{\boldmath $u$}}
\newcommand{\RR}{\mbox{\boldmath $R$}}
\newcommand{\BB}{\mbox{\boldmath $B$}}
\newcommand{\TL}{\mbox{$\overline{T}_{L}$}}
\newcommand{\TLN}{\mbox{$\overline{T}_{LN}$}}
\newcommand{\TN}{\mbox{$\overline{T}_{N}$}}

\newcommand{\xs}{\mbox{$\x_0$}}

\newcommand{\ys}{\mbox{$\y_0$}}

\newcommand{\rs}{\mbox{$\rr_0$}}

\title[Fluctuation dynamos at finite $\tau$]{Fluctuation dynamo at finite correlation times using renewing flows}

\author[Pallavi Bhat and Kandaswamy Subramanian]
{P\ls A\ls L\ls L\ls A\ls V\ls I\ns B\ls H\ls A\ls T$^1$%
  \thanks{Email address for correspondence: palvi@iucaa.ernet.in},\break
\and K\ls A\ls N\ls D\ls A\ls S\ls W\ls A\ls M\ls Y\ns S\ls U\ls B\ls R\ls A\ls M\ls A\ls N\ls I\ls A\ls N$^1$}

\affiliation{
IUCAA, Post Bag 4, Ganeshkhind, Pune 411 007, India.
}

\pubyear{2010}
\volume{650}
\pagerange{119--126}
\date{\today}

\begin{document}

\maketitle

\begin{abstract}
Fluctuation dynamos are generic to turbulent 
astrophysical systems. The only analytical model of the 
fluctuation dynamo, due to Kazantsev, assumes the velocity to be 
delta-correlated in time. This assumption breaks down for any 
realistic turbulent flow. We generalize the analytic model 
of fluctuation dynamo 
to include the effects of a
finite correlation time, $\tau$, using renewing flows. 
The generalized evolution equation for the longitudinal 
correlation function $M_L$ leads to the standard Kazantsev 
equation in the $\tau \to 0$ limit, and extends it to
the next order in $\tau$. We find that this evolution equation
involves also third and fourth spatial derivatives of $M_L$, 
indicating that the evolution for finite $\tau$ will be non-local
in general. In the perturbative case of small-$\tau$
(or small Strouhl number), it 
can be recast using the Landau-Lifschitz approach, to one
with at most second derivatives of $M_L$.
Using both a scaling solution and the WKBJ approximation, we
show that the dynamo growth rate is reduced when the
correlation time is finite. Interestingly, to leading
order in $\tau$, we show that the magnetic power spectrum,
preserves the Kazantsev form, $M(k) \propto k^{3/2}$,
in the large $k$ limit, independent of $\tau$.
\end{abstract}


\section{Introduction}

The continued existence of magnetic fields in most astrophysical
systems is thought to be due to dynamo action which converts kinetic
energy of the plasma into magnetic energy. In particular, fluctuation dynamos
are generic, and operate with minimal requirements of the underlying fluid flow.
A random flow with modest magnetic Reynolds number $\Rm\sim 100$
is sufficient to activate the fluctuation dynamo.
Here $\Rm=u/(q\eta)$ with $u$ and $q$ respectively
characteristic velocity and wavenumber 
of the flow and $\eta$ is the resistivity.
Hence fluctuation dynamos are considered to be ubiquitous in 
all astrophysical plasmas.

The analytical theory for the fluctuation dynamo was given by 
\citet{Kaz68}. A dynamical equation for the two point magnetic 
correlator was derived by using a simple model for the velocity field which 
is delta-correlated in time. This assumption of delta-correlation allows
one to convert the stochastic induction equation for the magnetic field
to a partial differential equation in real space for the 
longitudinal magnetic correlation function $M_L(r,t)$.
Its solution clearly showed for the first time that a random flow
with modest $\Rm$ can lead to the growth of the field.
Kazantsev then also predicted that the magnetic power spectrum
for a single scale or a large $\Pm$ turbulent flow,
scales asymptotically as
$M(k) \propto k^{3/2}$, for $q \ll k \ll k_\eta$, with $k_\eta$,
the wavenumber where resistive dissipation becomes important.
This spectrum is known as the Kazantsev spectrum.

Following the seminal work of \cite{Kaz68}, there has been considerable
interest in fluctuation dynamos, in terms of theoretical developments,
in terms of their direct simulation and in terms of various astrophysical 
applications \citep{MRS85,ZRS,KA92,KS97,RK97,
KS99,CFKV99,HBD04,Schek04,Scheketal05,BS05,SSH06,EV06,CVBLR09,MB10,
federrarth11,TCB11,Suretal12,SSBK12,Beresnyak12,BSS12,BS13}.
These works have clearly demonstrated that random (or turbulent) 
flows in a conducting plasma, with $\Rm > \Rmc \sim 30-500$, 
leads to the amplification of magnetic fields on the fast eddy turn over
time scale, usually much smaller than the age of the
astrophysical system. 
The $\Rmc$ depends on $\Pm = \nu/\eta$, where $\nu$ is the
viscosity and could even depend on the forcing wavenumber \citep{SB14}.
This rapid growth implies that fluctuation dynamos 
are crucial for the early generation of magnetic fields 
in primordial stars, galaxies and galaxy-clusters. 
It is therefore important to obtain a 
clear understanding of the fluctuation dynamo. 

Note that the feature of delta-correlation in time, assumed by \citet{Kaz68}, 
is not realistic in turbulent astrophysical plasmas. There the correlation
time, $\tau$, is expected to be at least of the order of the smallest eddy turn over time.
Thus, its important to understand the effects of finite
time correlation on the fluctuation dynamo.
This is the main motivation of the present work.

The effect of having a finite-$\tau$ on the magnetic energy 
growth has been considered by \citet{C97}, while \citet{SK01} examined its
consequences for the single point PDF in the ideal limit.
The correction to the evolution of the two point correlator due to 
having a finite-$\tau$ was considered by \citet{KRS02}; they 
however seem to have kept only a subset of the terms we derive here.
It was shown by \citet{MMBC11} that 
the results from simulations involving finite-$\tau$ velocity flows,
can be matched to the predictions using the Kazantsev equation 
provided the diffusivity spectrum is appropriately 
filtered out at small-scales. An analytic understanding of
the magnetic spectrum at finite-$\tau$ is however still lacking.

The present work uses random flows with finite time correlation known 
as renewing (or renovating) flows to develop an analytic 
generalization of the results of \citet{Kaz68} 
to include the effects of a finite correlation time.
\citet{Zeld88b} had used renewing flows for studying
the diffusion of scalars and the generation of vectors in random flows.
Such flows have also been used to study the effect of finite
correlation time on mean field dynamos \citep{DMSR84,GB92,KSS12}.
In an earlier letter \citep{BS14} (hereafter BS14), 
we gave a brief account of the our work on fluctuation dynamos 
using renewing flows, emphasizing  
an intriguing result that the Kazantsev spectrum
is in fact preserved even for such finite-$\tau$.
In the present paper, we present our detailed derivations
of the generalized Kazantsev equation and the results in BS14, 
as well as some new WKBJ analysis.
In the next section, we formulate the basic problem of fluctuation dynamos
in renewing flows. The detailed derivation of the evolution equation
for $M_L(r,t)$ which incorporates finite-$\tau$ effects, to the leading
order is given in section 3. Scaling and WKBJ analysis of this generalized
evolution equation is taken up in section 4, and we end with a discussion
of our results.

\section{Fluctuation dynamo in renewing flows}

The evolution of magnetic field, in a conducting fluid
with velocity $\uu$, is given by the induction equation,
\begin{equation}
\frac{\partial \BB}{\partial t} \;=\; \nabla \times \left( \uu \times \BB - \eta \nabla \times \BB \right).
\label{inductioneqn}
\end{equation}
The velocity field is a random flow which renews
itself every time interval $\tau$ \citep{DMSR84,GB92} and was given 
by \citet{GB92}(GB) as, 
\begin{equation}
\uu(\x)=\aaa\sin(\q\cdot\x+\psi),
\label{uturbdef}
\end{equation}
with $\aaa\cdot\q=0$ for an incompressible flow.
In each time interval $\left[(n-1)\tau, n\tau\right]$,\\ 
(i) $\psi$ is chosen uniformly random between 0 to $2\pi\,$; \\
(ii) $\q$ is uniformly distributed 
on a sphere of radius $q=\vert\q\vert$; \\
(iii) for every fixed $\hat{\q}=\q/q$, the direction of 
$\aaa$ is uniformly distributed 
in the plane perpendicular to $\q$. \\
Specifically, for computational ease, 
we modify the GB ensemble and use,
\be
a_i = \PPP{ij} A_j, ~~\PPP{ij}(\hat{\q}) = \delta_{ij} - \hat{q}_i\hat{q}_j
\label{aEns}
\ee
where $\AAA$ is uniformly distributed on a sphere of radius $A$, and
projects $\AAA$ to the plane perpendicular to $\q$.
Then on averaging over $a_i$ and using the fact that $\AAA$
is independent of $\q$, we have $\bbra{u}=0$ and,
\bea
\bra{a_ia_l}=\bra{a^2}\frac{\delta_{il}}{3}=\bbra{A_jA_k\PPP{ij}\PPP{lk}}&=&A^2\frac{\delta_{jk}}{3}\bbra{\PPP{ij}\PPP{lk}} 
=\frac{A^2}{3}\bbra{\PPP{il}}=\frac{2A^2}{3}\frac{\delta_{il}}{3}\nonumber\\
&\Rightarrow& \bra{a^2} = 2A^2/3
\label{aArel}
\eea
This modification in ensemble
does not affect any result using the renewing flows.
Condition (i) on $\psi$ ensures statistical homogeneity,
while (ii) and (iii) ensure statistical isotropy
of the flow.

The magnetic field evolution
in any time interval $\left[(n-1)\tau, n\tau\right]$ is
\be
B_i(\x,n\tau) \;=\; \int \mathcal G_{ij}(\x,\xs)
B_j( {\bf x_0},(n-1)\tau) \ d^3\xs
\ee
where $\mathcal G_{ij}(\x,\xs)$ is the Green's function of 
Eq.~(\ref{inductioneqn}).
S: added below
We define the magnetic two-point spatial correlation function as
\be
\bra{B_j(\x,t)B_l(\y,t)} = M_{jl}(r,t), \quad {\rm where}
\quad r=\vert\rr\vert = \vert(\x -\y)\vert,
\label{magcor}
\ee
and $\bbra{.}$ denotes an ensemble average.
Here we have assumed the statistical homogeneity and isotropy of the
magnetic field. Note that if the initial field is statistically 
homogeneous and isotropic, then this is preserved by the renewing
flow that we consider as we show explicitly below.
Then the evolution of the fluctuating field defined
by the two point correlation is,
\be
M_{ih}(\vert(\x-\y)\vert,n\tau)
= \int \bbra{\tilde{\mathcal{G}}_{ijhl}(\x,\xs,\y,\ys,\tau)}
M_{jl}(\vert(\xs-\ys)\vert,(n-1)\tau) \ d^3\xs \ d^3\ys.
\label{greenBB}
\ee
where $\bbra{.}$ around $\tilde{\mathcal{G}}$ 
denotes the average over the ensemble described above.
Here we could split the averaging on the right side of equation
between the Greens function and the initial magnetic correlator,
because the renewing nature of flow implies that the Greens function in the
current interval is uncorrelated to the magnetic 
correlator from the previous interval. 
The renewing nature of the flow also implies that $\tilde{\mathcal{G}}$ 
depends only on the time difference $\tau$ and not separately 
on the initial and final times in the interval $[(n-1)\tau,n\tau]$. 

To obtain $\tilde{\mathcal{G}}_{ij}(\x,\xs,\y,\ys,\tau)$ in the
renewing flow, we use the method introduced by GB. 
The renewal time, $\tau$, is split into two 
equal sub-intervals. In the first sub-interval $\tau/2$,  
resistivity is neglected and the frozen field is advected with twice
the original velocity. In the second sub-interval, $\uu$ is neglected and the field
diffuses with twice the resistivity. 
This method, plausible in the $\tau\to 0$ limit,
has been used to recover the standard mean field dynamo equations in renewing
flows \citep{GB92,KSS12}.

From the advective part of Eq.~(\ref{inductioneqn}), 
we obtain the standard Cauchy solution, 
in the first sub-interval $\tau/2=t_1-t_0$, 
\be
B_i(\x,t_1) = \frac{\partial x_i}{\partial x_{0j}}B_j( \xs,t_0)
\equiv J_{ij}(\x(\xs)) B_j( \xs,t_0).
\ee
Here $B_j( {\bf x}_0,t_0)$ is the initial field, 
which propagates
from $\xs$ at time $t_0$, to $\x$ at time $t_1 = t_0 +\tau/2$.  
In Eq.~(\ref{uturbdef}), the phase $\Phi = \q\cdot\x +\psi$
is constant in time as $d\Phi/dt = \q\cdot\uu =0$, from incompressibility. 
Then at time $t_1 = t_0 +\tau/2$, we integrate $d\x/dt = 2\uu$ to obtain,
\be
\x = \xs +\tau \uu = \xs + \tau {\bf a} \sin(\q\cdot\xs + \psi).
\label{traj}
\ee
Thus the Jacobian is,
\be
J_{ij}(\x(\xs)) = \delta_{ij} + \tau a_i q_j \cos(\q\cdot\xs + \psi).
\label{jacob}
\ee
It will be more convenient to work with 
the resulting field in Fourier space, 
\be
\hat{B}_i({\bf k},t_1) = \int J_{ij}({\bf x}({\bf x_0})) B_j( {\bf x_0},t_0) e^{-i \kk\cdot\x } d^3 \x.
\label{adveceq}
\ee
Then in the second sub-interval ($t_1,t=t_1+\tau/2$), 
only diffusion operates with resistivity $2\eta$ to give,
\be
\hat{B}_i(\kk,t) = G^{\eta}(\kk,\tau)\hat{B}_i( \kk,t_1) 
= e^{-(\eta \tau \kk^2)} \hat{B}_i( \kk,t_1),
\label{diffeq}
\ee
where $G^{\eta}$ is the 
resistive Greens function.
We combine \Eq{adveceq} and \Eq{diffeq}
to derive the evolution equation for the 
magnetic two point correlation function,
\be
\bra{\hat{B}_i(\kk, t)\hat{B}^*_h(\pp, t)} 
= e^{-\eta\tau(\kk^2+\pp^2)}\int \bbra{J_{ij}(\xs)J_{hl}(\ys) 
 e^{-i(\kk\cdot\x-\pp\cdot\y)}} M_{jl}(\rs,t_0) d^3\x d^3\y.
\label{corrlmaineq1}
\ee
The statistical homogeneity of the field
also implies the two-point magnetic correlator in Fourier space 
will be given by,
\be
\bra{\hat{B}_i(\kk,t)\hat{B}^*_h(\pp,t)}=(2\pi)^3 \delta^3(\kk-\pp) 
\hat{M}_{ih}(\pp,t).
\label{fouriercor}
\ee

We use Eq.~(\ref{traj}) 
to transform 
from $(\x,\y)$ to $(\xs,\ys)$ in Eq.~(\ref{corrlmaineq1}). 
Due to incompressibility of the flow,
the Jacobian of this transformation 
is unity.
We also write 
$\kk\cdot\xs - \pp\cdot\ys = \kk\cdot\rs + \ys\cdot(\kk-\pp)$ 
in Eq.~(\ref{corrlmaineq1}), transform from $(\xs,\ys)$ 
to a new set of variables $(\rs,\ys'=\ys)$, 
and integrate over $\ys'$. This leads to a delta
function in $(\kk-\pp)$ and  Eq.~(\ref{corrlmaineq1}) becomes,
\bea
&&\hat{M}_{ih}(\pp, t) =  
e^{-2\eta\tau\pp^2}\int \bra{R_{ijhl}} 
M_{jl}(\rs,t_0)e^{-i\pp\cdot\rs}d^3\rs \nonumber \\ 
&&\bra{R_{ijhl}}
=\bbra{J_{ij}(\xs)J_{hl}(\ys) 
e^{-i\tau(\aaa\cdot\pp) (\sin{A} - \sin{B})}} 
\label{avterm}
\eea
where, $A=(\xs\cdot\q+\psi)$ and $B=(\ys\cdot\q+\psi)$.
Due to statistical homogeneity of the renewing flow, we expect
$\bra{R_{ijhl}}$ to be only a function
of $\rs$, 
which we will see explicitly later.

\section{The generalized Kazantsev equation}
 
Exact evaluation of $\bra{R_{ijhl}}$ is difficult. 
However, we can motivate a 
Taylor series expansion of the exponential in  $\bra{R_{ijhl}}$ for small
Strouhl number $St = q \vert\aaa\vert \tau = q a\tau$, as follows.
Firstly in the argument of the exponential, $(\sin{A} - \sin{B}) 
= \sin(\q\cdot\rs/2) \cos(\psi + \q\cdot\RR_0)$,
where $\RR_0 = (\xs+\ys)/2$. Also for 
the kinematic fluctuation dynamo,
the magnetic correlation function peaks around the resistive scale 
$r_0 =\vert\rs\vert \sim 1/(q\Rm^{1/2})$,
or the spectrum peaks around $p \sim (q\Rm^{1/2})$. (Here
$p = \vert\pp\vert$.) Also $\Rm \sim a/(q\eta) \gg 1$. 
Thus, $q r_0 \ll 1$ and
$\sin(\q\cdot\rs) \sim \q\cdot\rs$. 
Subsequently the phase of the
exponential in Eq.~(\ref{avterm}) is of order 
$(p a \tau q r_0) \sim q a \tau = St$.
Thus for $St \ll 1$, one can expand the exponential
in Eq.~(\ref{avterm}) in $\tau$. We do this 
retaining terms up to $\tau^4$ order; keeping up to 
$\tau^2$ terms in Eq.~(\ref{avterm}), gives the Kazantsev equation, while the $\tau^4$ terms
give finite-$\tau$ corrections.
We get, 
\be
\bra{R_{ijhl}}=\bbra{H_{ijhl}[1 - i\tau\sigma - \frac{\tau^2\sigma^2}{2!} + \frac{i\tau^3\sigma^3}{3!} + \frac{\tau^4\sigma^4}{4!}] }, 
\label{expterms}
\ee
where $\sigma=(\aaa\cdot\pp)(\sin{A} - \sin{B})$ and
$H_{ijhl} = J_{ij}(\xs)J_{hl}(\ys)$ contains terms up to 
order $\tau^2$.
We note that \citet{KRS02} 
seem to have kept only up to $p^2$ terms in \Eq{expterms}.
\subsection{Kazantsev equation from terms up to order $\tau^2$}
We now consider all terms in \Eq{expterms} one by one up to the order $\tau^2$
and average over $\psi$, $\hat{\bf a}$ and $\hat{\bf q}$.
First consider $\bra{H_{ijhl}}$ from \Eq{expterms},
\be
\bra{H_{ijhl}} =\bbra{\delta_{ij}\delta_{hl}+\delta_{ij}a_h q_l \cos{A} +
\delta_{hl}a_i q_j \cos{B} 
+ a_i a_h q_j q_l  \frac{\tau^2}{2}\left(\cos(\q\cdot\rs)+\cos(2\q\cdot\RR_0+2\psi)\right)} 
\label{Hijhl1}
\ee
In \Eq{Hijhl1}, the second, third and last term on the right are
proportional to $\cos(...+n\psi)$ and hence go to
zero on averaging over $\psi$. Survival of such terms which 
depend explicitly on $\xs$, $\ys$ or $\RR_0$ and
would break statistical homogeneity.
The resulting expression after averaging over $\psi$ is,
\be
\bra{H_{ijhl}} = \bbra{\delta_{ij}\delta_{hl}+
a_i a_h q_j q_l  \frac{\tau^2}{2}\cos(\q\cdot\rs)} 
= \delta_{ij}\delta_{hl}
- \frac{\tau^2}{2}\partial_j\partial_l\bbra{a_i a_h \cos(\q\cdot\rs)} 
\label{Hijhl2}
\ee
where we have expressed $q_j\cos(\q\cdot\rs)$ as
$\partial_j\sin(\q\cdot\rs)$.
We find that the expression in \Eq{Hijhl2} contains
the two point velocity correlator or 
the turbulent diffusion tensor, given by,
\be
T_{ih} = \bra{u_i(\xs)u_h(\ys)} 
=\frac{\tau}{2} \bra{a_i a_h \sin(A) \sin(B)}=
 \frac{\tau}{4}\bra{a_i a_h \cos(\q\cdot\rs)}.
\label{difften}
\ee
Then we can express \Eq{Hijhl2} as,
\be
\bra{H_{ijhl}} = \delta_{ij}\delta_{hl}
- 2\tau\partial_j\partial_l T_{ih}. 
\label{part1order2}
\ee
Consider now the second term in \Eq{expterms},
$i\tau\bbra{H_{ijhl}~\sigma }$. We average over
$\psi$ and obtain statistically homogeneous terms,
\bea
\bbra{i\tau H_{ijhl}~\sigma} =&& \frac{i\tau^2}{2}\bbra{\aaa\cdot\pp~
\left[\delta_{ij}~a_h q_l~\sin(\q\cdot\rs) + \delta_{hl}~a_i q_j~\sin(\q\cdot\rs)\right]}\nonumber\\ 
=&& \frac{-i\tau^2}{2}~p_m\left[\delta_{ij}\partial_l\bbra{a_h a_m\cos(\q\cdot\rs)}
+\delta_{hl}\partial_j\bbra{a_i a_m\cos(\q\cdot\rs)}\right]\nonumber \\ 
=&& -2i\tau p_m  ~\left[\delta_{ij}~\partial_l T_{hm}+ \delta_{hl}~\partial_j T_{im}\right]
\label{part2order2}
\eea
where again in the last equation we have identified
and expressed in terms of the turbulent diffusion tensor.
Similarly for the third term in \Eq{expterms} to order $\tau^2$,
\bea
\bbra{H_{ijhl}\frac{\tau^2\sigma^2}{2}} =&& \frac{\tau^2}{2}\delta_{ij}\delta_{hl}p_m p_n
\bbra{a_m a_n \left[1-\cos(\q\cdot\rs)\right]} \nonumber \\
=&& 2\tau\delta_{ij}\delta_{hl}p_m p_n\left[T_{mn}(0) - T_{mn}\right].
\label{part3order2}
\eea
Now collecting all the simplified expressions of terms
in \Eq{expterms} up to order $\tau^2$, as given in 
\Eq{part1order2}, \Eq{part2order2} and \Eq{part3order2}, we
obtain,
\bea
\bbra{R_{ijhl}}=&& \delta_{ij}\delta_{hl} - 2\tau\partial_j\partial_l T_{ih}
 -i2\tau p_m  \left[\delta_{ij}\partial_l T_{hm}+ \delta_{hl}\partial_j T_{im}\right] \nonumber \\
+&& 2\tau\delta_{ij}\delta_{hl}p_m p_n\left[T_{mn}(0) - T_{mn}\right]
\label{simpR}
\eea
We then substitute \Eq{simpR} into \Eq{avterm}
and take the inverse Fourier transform of $\hat{M}_{ih}(\pp, t)$.
\be
{M}_{ih}(\rr, t) = \int  
(1-2\eta\tau\pp^2) \bra{R_{ijhl}} 
M_{jl}(\rs,t_0)e^{-i\pp\cdot(\rr-\rs)}d^3\rs \frac{d^3\pp}{(2\pi)^3}
\label{invFT}
\ee
where we also expand the exponential in the resistive Greens function
and consider only leading order term in $\eta$, relevant 
in the independent small $\eta$ (or $\Rm \gg 1$) limit.
In \Eq{invFT}, we consider only the first term in $\bbra{R_{ijhl}}$, 
$\delta_{ij}\delta_{hl}$ to multiply with $2\eta\tau\pp^2$
since all the other terms will be of the order higher than $\tau^2$.
In the case of the first two terms in $\bbra{R_{ijhl}}$ multiplying 
with unity, the integral in \Eq{invFT} is trivial
 with integration 
over $\pp$ first giving a delta function $\delta^3(\rr-\rs)$ which then leads
to all functions of $\rs$ simply turning into functions of $\rr$,
on integrating over $\rs$.
The other terms containing $p_i$ can be first written as 
derivatives with respect to $r_i$. 
For example, consider the integral in \Eq{invFT} containing the third term in 
$\bbra{R_{ijhl}}$, 
\bea
&&\int 2\tau~\delta_{ij}\delta_{hl}~p_m p_n~\left[T_{mn}(0) - T_{mn}\right]
M_{jl}(\rs,t_0)~e^{-i\pp\cdot(\rr-\rs)}~d^3\rs \frac{d^3\pp}{(2\pi)^3}\nonumber \\
=&&\int 2\tau (\frac{-\partial_m}{i})(\frac{-\partial_n}{i})\left[T_{mn}(0) - T_{mn}\right]
M_{ih}(\rs,t_0)~e^{-i\pp\cdot(\rr-\rs)}~d^3\rs \frac{d^3\pp}{(2\pi)^3}\nonumber \\
=&&-2\tau~\partial_m\partial_n\left[\left(T_L(0)-T_{mn}\right)M_{ih}(\rs,t_0)\right]
\label{pmul}
\eea
where we have used the fact that for a statistically 
homogeneous, isotropic and non helical velocity field, the correlation function
\be
T_{ih} = \left(\delta_{ih} -\rvu{i}\rvu{h}\right) T_{\rm N}(r,t)
+\rvu{i}\rvu{h} T_{\rm L}(r,t)
\ee
where $\hat{r_i}=r_i/r$
and hence
$T_{mn}(0)=\delta_{mn}T_L(0)$.
Here 
$T_{L}(r,t) =\rvu{i}\rvu{h} T_{ih}$ 
and $T_{N}(r,t) = (1/2r) [\partial (r^2 T_L)/\partial r]$ are, 
respectively, the longitudinal
and transversal correlation functions of the velocity field.
Carrying out all the steps, and noting that  
$(M_{ih}(\rr,t) -  M_{ih}(\rr,t_0))/\tau=\partial M_{ih}/\partial t$
in the limit $\tau \to 0$,
the resulting equation for $M_{ih}$ is given by,
\bea
\frac{\partial M_{ih}(\rr,t)}{\partial t} =&& 2\left(-[T_{ih} M_{jl}]_{,jl} + [T_{mh} M_{il}]_{,ml} 
+ [T_{im} M_{jh}]_{,jm} - [T_{mn} M_{ih}]_{,mn} \right) \nonumber \\
&&+ (2T_L(0) + 2\eta) ~~\nabla^2 M_{ih}
\label{finalKazcor}
\eea
Note that we have statistically homogeneous, isotropic and non helical
magnetic field, and hence similar to the velocity
correlation function, we have
$M_{ih} = \left(\delta_{ih} -\rvu{i}\rvu{h}\right) M_{\rm N}(r,t)
+\rvu{i}\rvu{h} M_{\rm L}(r,t)$.
Here 
$M_{L}(r,t)$ and $M_{N}(r,t)$ are, 
the longitudinal and transversal correlation functions of the magnetic field.
Then on contracting Eq.~(\ref{finalKazcor}) with $\rvu{i}\rvu{h}$
we obtain the dynamical equation for $M_L(r,t)$, the Kazantsev equation.
Note that we haven't yet performed averages over $\aaa$ and $\q$ because
we have simply identified the two point velocity correlator from \Eq{difften}
in expressions evaluated after averaging over $\psi$(as in \Eeqs{part1order2}{part2order2}{part3order2}).
We will explicitly have to perform the averages over $\aaa$ and $\q$ later
when we obtain the dynamical equation for $M_L$.\\
\subsection{Extending Kazantsev equation to higher order in $\tau$}

Next, we will consider the terms higher order in $\tau$, starting with
$\tau^3$ and then $\tau^4$. Interestingly, it turns out that all the terms of order $\tau^3$
go to $0$ on averaging. For example, from the second term in \Eq{expterms},
we obtain, $\tau^3\bbra{i(\pp\cdot\aaa)a_i a_h q_j q_l\cos{A}\cos{B}(\sin{A} - \sin{B})}$.
Here $\cos{A}\cos{B}\sin{A}=(1/2)\sin{(2A)}\cos{B}=(1/4)\left[\sin{(2A+B)}-\sin{(2A-B)}\right]$,
contain $\psi$ in their argument and hence go to $0$ on averaging.

Now we consider the terms of order $\tau^4$. The first contribution is from the third term in
\Eq{expterms}, $\tau^4\bbra{-[(\pp\cdot\aaa)^2/2]~ a_i a_h q_j q_l~\cos{A}\cos{B}[\sin{A} - \sin{B}]^2}$. 
On averaging over $\psi$, we obtain,
\bea
&-&\frac{\tau^4}{8}\bbra{a_i a_h~ (a_n p_n a_m p_m)~ q_j q_l \left[{\cos{(\q\cdot\rs)}-\cos{(2\q\cdot\rs)}}\right]} \nonumber \\
&=&\frac{\tau^4}{8}\bbra{a_i a_h~ (a_n p_n a_m p_m) ~\partial_j \partial_l \left[\cos{(\q\cdot\rs)}-\frac{\cos{(2\q\cdot\rs)}}{4}\right]}
\label{4p1}
\eea
We identify the terms in \Eq{4p1} with fourth order two point velocity correlators.
Three of such velocity correlators can be defined,
\bea
T_{mnih}^{x^2y^2}&=&\tau^2\bra{u_m(\x)u_n(\y)u_i(\x)u_h(\y)}, \nonumber \\
T_{mnih}^{x^3y}&=&\tau^2\bra{u_m(\x)u_n(\x)u_i(\x)u_h(\y)}, \nonumber \\
T_{mnih}^{x^4}&=&\tau^2\bra{u_m(\x)u_n(\x)u_i(\x)u_h(\x)}.
\label{Tmnih}
\eea
Again we multiply 
the fourth order velocity correlators
by $\tau^2$, as we envisage that 
$T_{ijkl}$ 
will be finite even in the $\tau\to 0$ limit,
behaving like products of turbulent diffusion.
Note that the renewing flow is not Gaussian random,
and hence higher order correlators of $\uu$ are not
the product of 
two-point correlators.
We consider the $\psi$ averaging of the velocity correlators in \Eq{Tmnih},
to obtain,
\bea
T_{mnih}^{x^2y^2}
&=&\tau^2\bbra{a_m a_n a_i a_h \sin^2{A}\sin^2{B}} 
=\frac{\tau^2}{4}\bbra{a_m a_n a_i a_h \left(1+\frac{\cos(2\q\cdot\rs)}{2}\right)}\\
\label{Tmnih1}
T_{mnih}^{x^3y}
&=&\tau^2\bbra{a_m a_n a_i a_h \sin^3{A}\sin{B}} 
=\frac{3\tau^2}{8}\bbra{a_m a_n a_i a_h \cos(\q\cdot\rs)}\\
\label{Tmnih2}
T_{mnih}^{x^4}
&=&\tau^2\bbra{a_m a_n a_i a_h \sin^4{A}}
=\frac{3\tau^2}{8}\bbra{a_m a_n a_i a_h }
\label{Tmnih3}
\eea
Now we can rewrite \Eq{4p1}, by expressing it in terms of the 
velocity correlators we have obtained in \Eqs{Tmnih1}{Tmnih2}.
We have,
\be
-\tau^2 p_n p_m \partial_j \partial_l\left[\frac{T_{mnih}^{x^2y^2}}{4}-\frac{T_{mnih}^{x^3y}}{3}\right]
\label{4p11}
\ee
Note that the first term in \Eq{Tmnih1} does not survive due to
the derivatives in \Eq{4p11}.
Similarly from the fourth term in \Eq{expterms}, the contribution of the order $\tau^4$
is given by,
\bea
&&i\tau^4\frac{(\pp\cdot\aaa)^3}{6}\left[\delta_{ij}a_h q_l \cos{B} + \delta_{hl}a_i q_j \cos{A} \right](\sin{A}-\sin{B})^3 \nonumber \\
=&&i\frac{\tau^4}{8} p_n p_m p_r \left(\bbra{\delta_{ij}a_k a_n a_m a_r \partial_l\left[2\sin(\q\cdot\rs)-\frac{\sin(2\q\cdot\rs)}{2}\right]}\right) \nonumber\\
-&&\tau^2 2p_n p_m p_r \left(\delta_{ij} \partial_l\left[\frac{T_{mnih}^{x^2y^2}}{4}-\frac{T_{mnih}^{x^3y}}{3}\right]\right) 
\label{4p2}
\eea
where we have again expressed in terms of velocity correlators from \Eqs{Tmnih1}{Tmnih2}.
Lastly, from the fifth term in \Eq{expterms}, 
$\frac{\tau^4}{24}\delta_{ij}\delta_{hl}\bbra{\left(\pp\cdot\aaa\right)^4\left(\sin{A}-\sin{B}\right)^4}$
we have,
\bea
&&=\frac{\tau^4}{16}\delta_{ij}\delta_{hl}p_m p_n p_r p_s\bbra{a_m a_n a_r a_s\left(\frac{3}{2}-2\cos(\q\cdot\rs)+\frac{\cos(2\q\cdot\rs)}{2}\right)} \nonumber \\
&&=\tau^2\delta_{ij}\delta_{hl}p_m p_n p_r p_s\left[\frac{T_{mnih}^{x^2y^2}}{4}-\frac{T_{mnih}^{x^3y}}{3}+\frac{T_{mnih}^{x^4}}{12}\right]
\label{4p3}
\eea
We again find that the integrand
determining the magnetic spectral tensor $\hat{M}_{ih}(\pp,t)$, 
is of the form $G(\pp)F_{ih}(\rs,t_0)$, where $G(\pp)$ is a polynomial
up to second order in $p_i$.
We can perform a simple inverse Fourier transform of 
$\hat{M}_{ih}(\pp, t)$, in Eq.~(\ref{avterm}) back to 
configuration space and then magnetic field correlation function
is,
\be
M_{ih}(\rr,t)
=\int G(\pp) F_{ih}(\rs,t_0)e^{i\pp\cdot(\rr-\rs)}d^3\rs 
\frac{d^3\pp}{(2\pi)^3}.
\label{Mihreal}
\ee
The $p_i$ in $G(\pp)$ above
can be written as derivatives with respect to $r_i$. 
Then integral over $\pp$ simply gives a delta
function $\delta^3(\rr-\rs)$ and this makes the integral over
$\rs$ trivial.
This was explicitly demonstrated earlier in \Eq{pmul}.

We then divide all the three contributions of the order $\tau^4$
in \Eeqs{4p11}{4p2}{4p3} by $\tau$. From the remaining $\tau^3$,
$\tau^2$ is absorbed into $T_{ijkl}$, 
leaving one $\tau$ which is treated as a small effective finite time parameter. 
Resulting extended equation for $M_{ih}$ is given by, 
\bea
&&\frac{\partial M_{ih}}{\partial t} = 2\left(-[T_{ih} M_{jl}]_{,jl} + [T_{jh} M_{il}]_{,jl} 
+ [T_{il} M_{jh}]_{,jl} - [T_{jl} M_{ih}]_{,jl} \right) + (2T_L(0) + 2\eta) ~~\nabla^2 M_{ih} \nonumber \\
&&+ \tau \left( \left[ \tilde{T}_{mnih} M_{jl}\right]_{,mnjl} 
- 2\left[\tilde{T}_{mnrh} M_{il}\right]_{,mnrl} + \left[\left(\tilde{T}_{mnrs}
+ \frac{T_{mnrs}^{x^4}}{12}\right)M_{ih}\right]_{,mnrs} \right) 
\label{finalcor}
\eea
where $\tilde{T}_{mnih} = T_{mnih}^{x^2y^2}/{4} - T_{mnih}^{x^3y}/{3}$,
$T_L(r) = \rvu{i}\rvu{j} T_{ij}$ with $\rvu{i} = r_i/r$. 
The first line in Eq.~(\ref{finalcor}) contains the terms which
give the Kazantsev equation as in \Eq{finalKazcor}, while the second line
contains the finite-$\tau$ corrections. 
We write
these latter terms as fourth derivative of the combined velocity and
magnetic correlators; however as both the velocity and magnetic fields
are divergence free, each spatial derivative only acts on one or the other.

We then contract Eq.~(\ref{finalcor}) with $\rvu{i}\rvu{h}$
to obtain the dynamical equation for $M_L(r,t)$.
On such a contraction, the terms in the first line
lead to the original Kazantsev equation for $M_L$. 
In order to perform such a contraction, we need to know the 
explicit form of the fourth order velocity correlator, $\tilde{T}_{mnih}$.
Such a fourth order two point correlator for a homogeneous and isotropic
velocity field can be expressed as,
\be
T_{mnih}=\rvu{mnih}\TL + \PP{(mn}\PP{ih)}\TN  + \rvu{(mn}\PP{ih)}\TLN
\label{expandT}
\ee
where $\rvu{mn}=\rvu{m}\rvu{n}$ and similarly $\rvu{mnih}=\rvu{m}\rvu{n}\rvu{i}\rvu{h}$.
$\PP{mn}=\delta_{mn}-\rvu{mn}$ is the configuration space projection operator. 
\be
\TL=\rvu{mnih}\tilde{T}_{mnih}, ~\TLN=\rvu{mn}\PP{ih}~\tilde{T}_{mnih}, ~\TN=\PP{mn}\PP{ih}~\tilde{T}_{mnih}/16
\ee
Lastly, the brackets $()$ in the subscripts
of two second rank tensors, denotes addition of all terms with all of the different permutations 
of the four indices considered in pairs. 
We will henceforth refer to all the ten terms in \Eq{expandT},
$\rvu{mnih}$,$\PP{mn}\PP{ih}$(and two other terms with permutations of the indices),
$\rvu{(mn}\PP{ih}$(and five other terms with permutations of the indices) as the basis tensors
(Although not all of them are orthogonal to each other).
For a divergence free(or incompressible) velocity field, the different correlation functions, $\TL$,
$\TN$ and $\TLN$, are related as,
\be
\TLN=\frac{1}{6r}\frac{d(r^2\TL)}{dr},~~ \TLN=\TN+\frac{r}{4}\frac{d(\TN)}{dr} \\
\ee
Consider the contraction of $\rvu{ih}$
with the first term in second line in \Eq{finalcor},\\
$\rvu{ih}\left[\tilde{T}_{mnih}M_{jl}\right]_{mnjl}=\rvu{ih}\tilde{T}_{mnih,jl}M_{jl,mn}$.
Then we have,
\bea
\rvu{ih}\tilde{T}_{mnih,jl}M_{jl,mn}
= \frac{1}{r^2}\left(\left[\rv{ih}\tilde{T}_{mnih}\right]_{,jl}-\left[\delta_{ij}r_h\tilde{T}_{mnih}\right]_{,j}
-\left[\delta_{il}r_h\tilde{T}_{mnih}\right]_{,l}\right. && \nonumber\\
\left.-\left[\delta_{jh}r_i\tilde{T}_{mnih}\right]_{,j}
-\left[\delta_{hl}r_i\tilde{T}_{mnih}\right]_{,j} + (\delta_{ij}\delta_{hl}+ \delta_{il}\delta_{jh})\tilde{T}_{mnih} \right) M_{jl,mn}&&
\label{p1cont}
\eea
We obtain a fourth order tensor from $\rvu{ih}\tilde{T}_{mnih,jl}$ which multiplies
another fourth order tensor $M_{jl,mn}$. To make this computation tractable,
we construct a table where we list the
coefficients of all the basis tensors.
We provide such a table in the Appendix~\ref{app1}, (Table.~\ref{tab1}).
Similarly for the second term in second line in \Eq{finalcor},
\bea
\rvu{ih}\left[\tilde{T}_{mnrh}M_{il}\right]_{lmnr}=&&(\rvu{h}\tilde{T}_{mnrh,l})(\rvu{i}M_{il,mnr}) = 
\frac{1}{r^2}\left(\left[\rv{h}\tilde{T}_{mnrh}\right]_{,l}-\delta_{lh}\tilde{T}_{mnrh}\right) \times \nonumber \\
&&\left(\left[r_i M_{il}\right]_{,mnr} -\delta_{ir}M_{il,mn}-\delta_{in}M_{il,mr}-\delta_{im}M_{il,nr} \right)
\label{p2cont}
\eea 
Again we have given the expansion of the fourth order objects $(\rvu{h}\tilde{T}_{mnrh,l})$ and
$(\rvu{i}M_{il,mnr})$ (in terms of basis tensors),  in Table.~\ref{tab2} in the Appendix~\ref{app1}.
Then lastly we have the third term from the second line in \Eq{finalcor},
\bea
\rv{ih}\left[\tilde{T}_{mnrs}M_{ih}\right]_{,mnrs}&=&\tilde{T}_{mnrs}\left(\left[\rv{ih}M_{ih}\right]_{,mnrs}
- (r_{ih})_{,m}M_{ih,nrs} -(r_{ih})_{,n}M_{ih,mrs}\right. \nonumber \\
- (r_{ih})_{,r}M_{ih,nms}&-&(r_{ih})_{,s}M_{ih,nrm} - (r_{ih})_{mn}M_{ih,rs}
- (r_{ih})_{mr}M_{ih,ns}  \nonumber \\
- (r_{ih})_{ms}M_{ih,rn}&-& (r_{ih})_{ns}M_{ih,mr} - \left. (r_{ih})_{rs}M_{ih,mn}- (r_{ih})_{rn}M_{ih,ms}\right)
\label{p3cont}
\eea
Here the two fourth order tensor objects multiplying each other are $\tilde{T}_{mnrs}$ 
and $r_{il} M_{jl,mnrs}$ and the expansion of such fourth order objects 
in terms of basis tensors can be again found in
Table~\ref{tab3}, in the Appendix~\ref{app1}.

The tables \ref{tab1}, \ref{tab2} and \ref{tab3} are useful in making the algebra of
the all the fourth order terms in \Eeqs{p1cont}{p2cont}{p3cont} tractable.
In each of the tables, we list the expansion of all the individual fourth
order objects in terms of the basis tensors. The basis tensors form the rows, while
the expansion coefficients in \Eeqs{p1cont}{p2cont}{p3cont} are listed as columns.
Note that the first column is the list of the basis tensors. Then the subsequent
columns list the expansion coefficients (of the respective basis tensor) for
each fourth order terms in the \Eeqs{p1cont}{p2cont}{p3cont}.
Then we sum the contributions from each row, separately for the magnetic and velocity 
parts. The last but one column in Table~\ref{tab1} and the last columns in Table~\ref{tab2}
and Table~\ref{tab3} give the resulting sum divided by $r^2$.
We then finally multiply the sum obtained for the magnetic part
with the sum from the velocity part. 

Here, we note that when we multiply
one group of the basis tensors with another, all of them goto zero,
but yield a constant when multiplied within the same group.
For example product of $\rvu{mnih}$ and $\rvu{mn}\PP{ih}$ goes to zero,
but product of $\rvu{mnih}$ with itself naturally produces unity. Then the product
of $\rvu{mn}\PP{ih}$ with $\rvu{ih}\PP{mn}$ (or the other four similar kind of terms)
goes to 0, but with itself gives a value of 2.
Lastly, product of $\PP{mn}\PP{ih}$ with $\PP{mi}\PP{nh}$ (or $\PP{ni}\PP{mh}$)
gives a value of 2, but with itself gives a value of 4.

By multiplying the velocity part with the magnetic part in
this manner, we finally obtain the additional terms from the contractions, due to finite $\tau$
and extend the Kazantsev equation to the form, 
\bea
&&\frac{\partial M_L(r,t)}{\partial t} = 
\frac{2}{r^4} \frac{\partial}{\partial r} \left( r^4 \eta_{tot} \frac{\partial M_L}{\partial r} \right) + G M_L \nonumber \\
&&+ \tau M_L^{''''} \left({\overline{T}_L} + \frac{\overline{T}_L(0)}{12}\right)
+ \tau M_L^{'''}\left(2 \overline{T}_L^{'} + \frac{8 \overline{T}_L}{r} 
+ \frac{2 \overline{T}_L(0)}{3 r} \right) \nonumber \\
&&+ \tau M_L^{''} \left(\frac{5 \overline{T}_L^{''}}{3} + \frac{11 \overline{T}_L^{'}}{r} 
+ \frac{8 \overline{T}_L}{r^2} + \frac{2 \overline{T}_L(0)}{3 r^2}\right)  \nonumber \\
&&+ \tau M_L^{'} \left(\frac{2 \overline{T}_L^{'''}}{3} + \frac{17 \overline{T}_L^{''}}{3r} 
+ \frac{5 \overline{T}_L^{'}}{r^2} - \frac{8 \overline{T}_L}{r^3} - 
\frac{2 \overline{T}_L(0)}{3 r^3} \right) 
\label{finMleq}
\eea
Here, 
$\eta_{tot} = \eta + T_L(0) - T_L(r)$ and 
$G = -2 \left(T_L^{''} + 4 T_L^{'}/r\right)$.
Here again the first line gives us the original Kazantsev equation and 
the rest of the terms form the extended part and have the parameter
$\tau$ multiplying them. We will refer to \Eq{finMleq} as the generalized Kazantsev equation
incorporating finite $\tau$ effects.
To proceed further, and solve the generalized Kazantsev equation \Eq{finMleq}
we need to firstly evaluate the second and fourth order velocity correlators explicitly 
for the renewing flow from \Eq{difften} and \Eeqs{Tmnih1}{Tmnih2}{Tmnih3} respectively.
Consider first the two point velocity correlator,
\be
T_{ij} =
\frac{\tau}{4}\bra{A_l A_m P_{il} P_{jm} \cos(\q\cdot\rr)}
=\frac{A^2\tau}{12}\bra{P_{ij}\cos(\q\cdot\rr)}
=\frac{a^2\tau}{8}\left[\delta_{ij} + \frac{1}{q^2}
\frac{\partial^2}{\partial r_{i}r_{j}}\right]
j_0(q r).
\label{Tij}
\ee
Here, we have made use of the results in \Eqs{aEns}{aArel}, i.e.
we have substituted for $\aaa$ in terms of ${\bf A}$,
and first averaged over ${\bf A}$. 
Similarly in the expression for $T_{mnih}^{x^2y^2}$ in \Eq{Tmnih1}, 
we substitute $a_m=A_s\PPP{ms}$, $a_n=A_t\PPP{nt}$, 
$a_i=A_u\PPP{iu}$ and $a_h=A_v\PPP{hv}$. 
Then we have,
\be
T_{mnih}^{x^2y^2}=\frac{\tau^2 A^4}{60}\bbra{\PPP{(mn}\PPP{ih)}\left(1+\cos(2\q\cdot\rr)\right)}.
\label{Tmnih11}
\ee
The first part in \Eq{Tmnih11} is evaluated to be 
$\bra{\PPP{(mn}\PPP{ih)}}=8/15\left(\delta_{(mn}\delta_{ih)}\right)$.
And the second part in \Eq{Tmnih11} is given as,
\bea
\bbra{\PPP{(mn}\PPP{ih)}\cos(2\q\cdot\rr)}=&&\left[\left(\delta_{mn}+\partial_m\partial_n\right)\left(\delta_{ih}+\partial_i\partial_h\right)
+\left(\delta_{mi}+\partial_m\partial_i\right)\left(\delta_{nh} +\partial_n\partial_h\right) \right.\nonumber \\
&& \left. +\left(\delta_{mh}+\partial_m\partial_h\right)\left(\delta_{in}+\partial_i\partial_n\right)\right] j_0(2qr_0) \nonumber
\eea
\bea
=&&-24\left(\frac{j_0(2z)}{(2z)^2}+\frac{3\partial_{2z}j_0(2z)}{(2z)^3}\right)\rvu{mnih} + \left(j_0 + \frac{2\partial_{2z}j_0(2z)}{2z} 
-\frac{3\partial_{2z}j_0(2z)}{(2z)^2}- \frac{9\partial_{2z}j_0(2z)}{(2z)^3}\right)\nonumber \\
&&\left[\PP{(mn}\PP{ih)}\right]
+ \left(-\frac{4\partial_{2z}j_0(2z)}{z} +\frac{12\partial_{2z}j_0(2z)}{(2z)^2}+\frac{36\partial_{2z}j_0(2z)}{(2z)^3}\right) 
\left[\rvu{(mn}\PP{ih)}\right]
\label{Tmnihx2y2}
\eea
where $z=qr$ and the derivative $\partial_{2z}$
is derivative with respect to $2z$.
We get a similar expression as in \Eq{Tmnihx2y2} also for 
$T_{mnih}^{x^3y}=\frac{A^4}{40}\bbra{\PPP{(mn}\PPP{ih)}\cos(\q\cdot\rr)}$,
with all the $(2z)$ replaced by $z$ and $\partial_{2z}$ by $\partial_{z}$.
We give the expressions for $\overline{T}_L^{x^2y^2}$ and $\overline{T}_L^{x^3y}$,
\be
\overline{T}_L^{x^2y^2}=
\frac{ -9a^4\tau^2}{10}\left(\frac{3\partial_{2z}j_0(2z)}{(2z)^3}+\frac{j_0(2z)}{(2z)^2}\right),~
\overline{T}_L^{x^3y}=
\frac{-27a^4\tau^2}{20}\left(\frac{3\partial_{z}j_0(z)}{z^3}+\frac{j_0(z)}{z^2}\right),
\label{longbarT}
\ee 
(The above expressions correct the missing $\sim a^4\tau^2$ factors in Eq.~(18) in BS14)
These latter equalities give 
the explicit expressions of these 
fourth order 
correlators for the renewing flow.
\Eq{finMleq} allows eigen-solutions of the form
$M_L(z,t)=\tilde{M}_L(z)e^{\gamma \tilde{t}}$, 
where $\tilde{t}=t\eta_t q^2$, with $\eta_t = T_L(0)= 
a^2\tau/12 = A^2\tau/18$,
and $\gamma$ is the growth rate.
Boundary conditions are given as
$M_L'(0,t) =0$, $M_L \to 0$ as $r\to \infty$. 
Implications of the higher spatial derivative terms
are discussed below.

\section{Growth rate and magnetic spectrum at finite correlation time}

We now discuss the solution of \Eq{finMleq} to examine the 
finite correlation time modification to the growth rate and 
magnetic correlation function or its energy spectrum.
For the latter, we focus particularly on the large $k$ (or small $r$) 
behaviour. Recall that in the $\tau \to 0$ limit the magnetic spectrum is
of the Kazantsev form, $M(k) \propto k^{3/2}$ for $q \ll k \ll k_\eta$.
Our aim is to determine how this gets modified in the presence of finite
correlation time effects. For this purpose, 
we employ two different approaches.
First, we recall in more detail the scaling solution discussed
in BS14. We also then present a WKBJ analysis to derive $M_L(r,t)$
in the small $r$ limit, and hence the magnetic spectrum. 

In both approaches, 
to derive the standard Kazantsev spectrum  
in the large $k$ limit, and its finite-$\tau$ modifications, 
it suffices to go to the limit of small $z=qr\ll1$.
Expanding the Bessel functions 
in \Eqs{Tij}{longbarT} in this limit, and substituting 
$M_L(z,t)=\tilde{M}_L(z)e^{\gamma \tilde{t}}$, 
\Eq{finMleq} becomes,
\bea
&&\gamma \tilde{M}_L(z)=
\left(\frac{2\eta}{\eta_t} + \frac{z^2}{5}\right)\tilde{M}_L^{''}
+\left(\frac{8\eta}{\eta_t} + \frac{6 z^2}{5}\right) \frac{\tilde{M}_L^{'}}{z} + 2\tilde{M}_L
\nonumber \\
&&+ \frac{9 \bar\tau}{175}\left(\frac{z^4}{2}\tilde{M}_L^{''''} 
+ 8z^3 \tilde{M}_L^{'''} 
+36 z^2  \tilde{M}_L^{''}
+48 z \tilde{M}_L^{'}\right)
\label{smallrEq}
\eea
where $\bar\tau = \tau \eta_t q^2 = (St)^2/12$ and prime is now z-derivative.

For the solution near the origin, where $z \ll \sqrt{\eta/\eta_t}$,
it suffices to approximate $\tilde{M}_L$ as a parabola and write 
$\tilde{M}_L(z) =M_0(1 - z^2/z_\eta^2)$. From
\Eq{smallrEq}, we find  
$z_\eta = qr_\eta=[240/(2-\gamma)]^{1/2} [\Rm (St)]^{-1/2}$.
The $\bar\tau$ dependent terms, which are small because both
$z$ and $\bar\tau$ are small, do not affect this result.
Thus for $\Rm\gg1$, the resistive scale $r_\eta \ll 1/q$ 
(or $k_\eta = 1/r_{\eta} \gg q$), although one has 
to go to sufficiently large $\Rm \gg 240/((2-\gamma)St)$ 
for this conclusion to obtain. 

In order to determine the magnetic correlation function 
for spatial scales larger than $z_\eta$, and also
obtain the growth rate,
we have to more fully analyze \Eq{smallrEq}. 
We see that this evolution equation (or \Eq{finMleq}), 
also has higher order (third and fourth) spatial derivatives  
when going to finite-$\tau$ case.
This indicates that for finite $\tau$, $M_L$ evolution is actually 
nonlocal, determined by an integral type equation; but
whose leading approximation for small $\bar\tau$ is the
local equation \Eq{smallrEq}. 
However these higher
derivative terms only appear as perturbative terms 
multiplied by the small parameter $\bar\tau$. Then it is
possible to use 
the Landau-Lifshitz type approximation,
earlier used in treating the effect of radiation reaction
force in electrodynamics (see \citet{LL} section 75).
In this treatment, one first ignores the perturbative terms proportional
to $\bar\tau$, which gives basically Kazantsev equation for 
$\tilde{M}_L$, and uses this 
to express $\tilde{M}_L^{'''}$ and $\tilde{M}_L^{''''}$ in terms
of the lower order derivatives $\tilde{M}_L^{''}$ and $\tilde{M}_L^{'}$.  

We will find that for both the scaling solution and for 
determining the asymptotic WKBJ 
solution, these higher order derivatives are only required in the limit
$z\gg z_\eta$. In this limit we have from \Eq{smallrEq} at the zeroth order in $\bar{\tau}$,
\be
\frac{z^2}{5}\tilde{M}_L^{''} = 
- \frac{6 z}{5}\tilde{M}_L^{'} + (\gamma-2)\tilde{M}_L
\label{zeroth}
\ee
Differentiating this expression first once and then twice gives,
\be 
z^3\tilde{M}_L^{'''} =  -8 z^2 \tilde{M}_L^{''} 
- z(16-5\gamma_0)\tilde{M}_L^{'}, \  
z^4\tilde{M}_L^{''''} = (56+5\gamma_0) 
z^2\tilde{M}_L^{''} +10(16-5\gamma_0) z\tilde{M}_L^{'}.
\label{higher}
\ee
Here $\gamma_0$ is the growth rate which obtains for the
Kazantsev equation in the $\tau\to 0$ limit.
We now turn to the scaling solution approach.

\subsection{Growth rate and magnetic correlations from a scaling solution}

Consider the solution for $z_\eta \ll z \ll 1$.
In this limit, ignoring terms depending on $\eta/\eta_t$, 
\Eq{smallrEq} itself is scale free, as 
scaling $z\to c z$ leaves it invariant. Thus
the resulting equation has power law solutions of the form
$\tilde{M}(z) = \bar{M}_0 z^{-\lambda}$. To find the form of this 
solution, we first substitute the expressions in \Eq{higher} back into 
the full \Eq{smallrEq}. We get after neglecting the $\eta/\eta_t$ terms,
\be
\tilde{M}_L^{''}z^2 \left( \bar{\tau}\gamma_0\frac{9}{70} + \frac{1}{5}\right) + \tilde{M}_L^{'}z \left( \bar{\tau}\gamma_0\frac{27}{35} + \frac{6}{5}\right)
+(2-\gamma)\tilde{M}_L=0
\label{smallrEq2}
\ee
We find the interesting result that the coefficients of the perturbative terms
in \Eq{smallrEq} 
are such that all perturbative terms which do not depend on 
$\gamma_0$ cancel out in \Eq{smallrEq2} !

As advertised \Eq{smallrEq2} admits power law solutions of the form
$\tilde{M}_L(z) = \bar{M}_0 z^{-\lambda}$, with $\lambda$ determined by,
\be
\lambda^2- 5\lambda + \frac{5(2-\gamma)}{1+\frac{9}{14}\gamma_0 \bar{\tau}} = 0;
\quad {\rm so} \ \lambda = \frac{5}{2} \pm i \lambda_I,~
\lambda_I = \frac{1}{2}\left[\frac{20(2-\gamma)}{(1 + 9\gamma_0\bar\tau/14)}-25\right]^{1/2}
\label{smallrEq3}
\ee
More important is the fact that the real part of $\lambda$ is $\lambda_R=5/2$,
independent of the value of $\bar{\tau}$!
We can also get the approximate growth rate 
assuming $\Rm\gg1$, following
an argument from \citet{GCS96}. 
These authors looked at \Eq{smallrEq3} as an equation for
$\gamma(\lambda)$ and argued that the growth rate is determined
by substituting in to \Eq{smallrEq3},
the value of $\lambda=\lambda_m$ where 
$d\gamma/d\lambda =0$. This gives 
\be
\gamma_0\approx 3/4, \quad {\rm and} \quad 
\gamma \approx (3/4)(1 - (45/56)\bar\tau).
\label{growths}
\ee
Note that \Eq{growths} also implies 
$\lambda_I \approx 0$. (Including the effects of resistivity
gives $\lambda_I$, a small 
positive
non zero value $\propto 1/(\ln(\Rm))$
as will be shown below).
The $\gamma_0$ we get matches with that of \citet{KA92},
obtained from the evolution equation of $M(k,t)$.
It is also important to note that the growth rate is reduced for a 
finite $\bar\tau$. This was found in 
simulations which directly compare with 
an equivalent Kazantsev model \citep{MMBC11}.

The form of the magnetic correlation $M_L$ for $z_\eta \ll z \ll1$ 
can also be found from \Eq{smallrEq3}. It is given by
\be
M_L(z,t) = e^{\gamma\tilde{t}} \tilde{M}_0
z^{-5/2} \sin\left(\lambda_I \ln(z) + \phi\right),
\label{finsoln}
\ee
where $\tilde{M}_0$ and $\phi$ are constants.
Thus in this range, $M_L$ varies dominantly as $z^{-5/2}$,
modulated by the weakly varying sine factor (as $\lambda_I$ is small).
We will use this below to determine the asymptotic magnetic spectrum.
Before that, we turn to the alternate approach to determining
$\gamma$ and $M_L$, using the WKBJ approximation, 
which also allows one to incorporate the effects of the small resistive terms.

\subsection{Growth rate and Magnetic correlation function using WKBJ analysis}

First it is convenient to define a scaled co-ordinate 
$\zr = (\sqrt{\eta_t/\eta}) \ z$.
In terms of this new coordinate the resistive scale will have $\zr\sim1$,
where as the forcing scale, $z=1$ corresponds to $\zr \sim \sqrt{\Rm} \gg1$.
Now substituting the expressions in \Eq{higher} back into 
the full \Eq{smallrEq} we get, 
\be
\frac{d^2\tilde{M}_L}{d\zr^2}
\left(2+ \bar{\tau}\gamma_0\frac{9\zr^2}{70} + \frac{\zr^2}{5}\right) + 
\frac{d\tilde{M}_L}{d\zr} \left(\frac{8}{\zr} 
+ \bar{\tau}\gamma_0\frac{27 \zr}{35} 
+ \frac{6\zr}{5}\right)
+(2-\gamma)\tilde{M}_L=0
\label{smallrEq2eta}
\ee
As remarked earlier, the coefficients of the perturbative terms
in \Eq{smallrEq} 
are such that all perturbative terms which do not depend on 
$\gamma_0$ cancel out in \Eq{smallrEq2eta}.

Further, in order to implement the boundary condition at $\zr=0$, 
under WKBJ approximation, it is better to transform to a new variable 
$x$, where $\zr=e^x$. Also to eliminate first derivative terms 
in the resulting equation we substitute $\tilde{M}_L(x) = g(x) W(x)$,
and choose $g(x)$ to satisfy the differential equation,
\be
\frac{1}{g}\frac{dg}{dx} =
-\frac{5}{2} \frac{\left(6 + \zr^2 F\right)} 
{\left(10 + \zr^2 F \right)}, \quad {\rm with} \quad
F = (1 + (9/14)\bar\tau \gamma_0).
\label{geq} 
\ee
Then $W$ satisfies,
\be
\frac{d^2W}{dx^2} + p(x) W = 0
\label{wkbeq}
\ee
where
\be
p(x) = \frac{A_0 \zr^4 -B_0 \zr^2 -225}
{\left(10 + F\zr^2 \right)^2},
\label{pofx}
\ee 
\be
A_0 =  5F \left(\frac{3}{4} -\frac{45}{56} \bar\tau\gamma_0 - \gamma\right), 
\quad
B_0 = 5 \left(10\gamma + \frac{171}{14} \bar\tau\gamma_0 -1\right).
\label{a0b0}
\ee
The WKBJ solutions to this equation are linear combinations of
\be
W = {1\over p^{1/4}} \exp (\pm i\int^x p^{1/2} dx)
\ee
Note that as $\zr \to 0$, $x\to -\infty$ and $p \to -9/4$; so the WKBJ 
solutions are in the form of growing and decaying exponentials at this end.
And as $\zr$ increases to a large enough value, $p(x)$ goes through
a zero at say $\zr=\zr_0$ (or $x=x_0$) 
and becomes positive for $\zr > \zr_0$.
The solution then becomes oscillatory. Note that at 
$\zr \to +\infty$, one would again want to solution to
decay, and so $p(x)$ should become negative. This cannot be
seen in \Eq{pofx}, as it is valid only for $z \ll 1$ (or 
$\zr \ll \sqrt{\Rm}$),
but would require one to consider \Eq{finMleq} in the opposite 
limit of $z\gg 1$ (or $\zr \gg \sqrt{\Rm}$).
In such a limit one has $T_L(r) \to 0$, $\overline{T}_L(r) \to 0$, and
again using the Landau-Lifshitz ansatz to eliminate $\tilde{M}_L^{''''}$,
$\tilde{M}_L(z)$ now satisfies
\be
\gamma \tilde{M}_L(z)=
\left(\frac{2\eta}{\eta_t} + 2 +\bar\tau\alpha\right)\tilde{M}_L^{''}
+8\left(\frac{\eta}{\eta_t} + 1\right) \frac{\tilde{M}_L^{'}}{z},
\label{largezeq}
\ee
where $\alpha = (q^2 \overline{T}_L(0)\gamma_0)/[12 (\eta+\eta_t)]$.
We can again transform to the $x$-coordinate, and write $\tilde{M}_L = gW$.
Then in this limit of $z \gg 1$, 
$W$ again satisfies \Eq{wkbeq} with now
\be
\frac{1}{g}\frac{dg}{dx} =
-e^x\frac{(1 + \eta_t/\eta)}{2(2 +2\eta_t/\eta + \bar\tau\alpha)},
\
p(x) = -e^{2x} \frac{(1 + \eta_t/\eta) + \gamma}
{(2 + 2\eta_t/\eta + \bar\tau\alpha)^2}.
\label{gandp}
\ee
We see that $p(x)$ is now negative definite and so again
one has exponentially damped solutions for $W$. Since $p(x) >0$
for $\zr> \zr_0$, and is negative at $\zr \gg \sqrt{\Rm}$, 
there would again be a point, say $\zr=\zr_c$ (or $x=x_c$), 
where it would go to zero.
We approximate our WKB treatment by assuming that \Eq{smallrEq2eta} is valid
for $z < 1$ and \Eq{largezeq} is valid for $z>1$. 
The outer transition point $\zr_c$
then can be taken to be the boundary between these two regions.
We will see that the $\zr_c$ dependence, in the determination of the growth rate
and $\tilde{M}_L$ only comes within a logarithm, and so our results
are not very sensitive to its exact value. This insensitivity
to the outer boundary condition has been remarked earlier by
several authors \citep{KA92,GCS96,SBK02,BS05}.

The requirement that the oscillatory solution in the region
$\zr_0 < \zr < \zr_c$ match on to the 
growing exponential near $\zr \ll  \zr_0$ and the decaying
exponential as $\zr \gg \zr_c$, gives the standard condition
\citep{BO78,MS91,KS97} 
on the the eigenvalue $\gamma$
\be
\int_{x_0}^{x_c} p^{1/2}(x) dx  = {(2n + 1)\pi \over 2} .
\label{wkb}
\ee
We will find that $\zr_0$ is large enough that one can neglect
the constant terms in \Eq{pofx}. Then the integral in \Eq{wkb} can
be done exactly and leads to the condition,
\be
A_0^{1/2}\left[ \ln\left( \frac{\zr_c}{\zr_0} + 
\left(\frac{\zr_c^2}{\zr_0^2}-1\right)^{1/2} \right) 
-\left(1-\frac{\zr_0^2}{\zr_c^2}\right)^{1/2}\right] 
=\frac{\pi F}{2}.
\label{eigen1}
\ee
Here we have taken $n=0$ which corresponds to the fastest growing
eigenfunction. We will also find self-consistently that for
large $\Rm$, $\zr_c^2/\zr_0^2 \gg1$. In this case 
\Eq{eigen1} gives for the growth rate, 
\be
\gamma = 
\frac{3}{4} -\frac{45}{56} \bar\tau\gamma_0 
-\frac{\pi^2}{5} \frac{(1 + (9/14)\bar\tau\gamma_0)}{(ln(2\zr_c/\zr_0))^2}.
\approx \frac{3}{4}\left[1 -\frac{45}{56} \bar\tau \right] 
-\frac{\pi^2}{5} \frac{(1 + (27/56)\bar\tau)}{(\ln(\Rm))^2}.
\label{growth}
\ee
In the latter part of \Eq{growth}, we have 
used self-consistent estimates of $\gamma_0\sim 3/4$,
$\zr_c \sim \sqrt{\eta_t/\eta} z_c \sim \sqrt{\Rm}$ and 
$\zr_0 \sim \sqrt{B_0/A_0} \sim \ln(\Rm)$, and so
also neglected $\ln\zr_0$ compared to $\ln\zr_c$.
This result for the growth rate exactly matches with that
obtained earlier by BS14 in the limit of large $\Rm$ using 
a scaling solution (see \Eq{growths} above). It of course corrects this
estimate for finite $\Rm$. We also see from \Eq{growth} that
the growth rate is insensitive (more correctly only logarithmically sensitive) 
to the exact value of $\zr_c$, the upper zero of $p(x)$.

The WKB analysis also gives the form of the eigenfunction between the
two zeros
\be
W(x) \approx \frac{1}{p^{1/4}} \sin\left[ \int^x_{x_1} (p)^{1/2} dx 
+ {\pi\over 4}\right] 
\approx
\frac{(\ln\Rm)^{1/2}}{\pi^{1/2}} 
\sin\left[ \frac{\pi}{\ln\Rm} \ln\left(\frac{\zr}{\zr_0}\right) 
+ {\pi\over 4}\right] 
\label{Woscil}
\ee
where for the latter expression we have taken the large $\zr > \zr_0\gg1$ limit
which is applicable here. Also for $\zr \gg 1$, we can see from \Eq{geq}
that $(1/g) (dg/dx) \to -5/2$ independent of the value of $\bar\tau$.
Thus in this limit $g(x) \propto \exp(-5x/2)$. Since $M_L(z) \propto 
e^{\gamma \tilde{t}} gW$, the WKB solution
for the region $z_\eta \ll z \ll 1$ is then given by,
\be
M_L(z,t) = e^{\gamma\tilde{t}} \tilde{M}_0
z^{-5/2} 
\sin\left[ \frac{\pi}{\ln\Rm} \ln\left(\frac{z}{z_0}\right)
+ {\pi\over 4}\right] 
\label{finsoln2}
\ee
This again matches with the result obtained from the scaling solution,
improving it by fixing the constants there, in particular $\lambda_I$.
We see that the dominant variation of  $M_L(z,t)$ in this regime is the
power law behaviour $M_L \propto z^{-5/2}$, modulated by the
weakly varying sine factor, as before.

The power law scaling of the magnetic correlation function can be
translated to the scaling of the magnetic power spectrum. It is
straightforward to show that 
the magnetic power spectrum is related to the longitudinal 
correlation function $M_L$ by (cf. \citet{BS00}),
\be
M(k,t) = \int dr (kr)^3 M_L(r,t) j_1(kr) 
\label{MkMl}
\ee
The spherical Bessel function $j_1(kr)$ is peaked around
$ k\sim 1/r$, and every value of $k$ in $M(k,t)$ gets
dominant contribution in the integral in \Eq{MkMl} 
from values of $r\sim 1/k$. Therefore a power law 
behaviour of 
$M_L \propto z^{-\lambda_R}$
for a range of $z_\eta \ll z=qr \ll1$,
translates into a power law
for the spectrum $M(k) \propto k^{\lambda_R -1}$ in the 
corresponding wavenumber range $q \ll k \ll q/z_\eta$. 
Both the scaling solution in \Eq{finsoln} and the WKBJ 
solution given in \Eq{finsoln2}, show that 
in the range $z_\eta \ll z \ll1$,
$M_L$ dominantly varies as a power law with $\lambda_R=5/2$,
independent of $\tau$. This then leads to the remarkable result
emphasized by BS14 that
the magnetic spectrum is of the Kazantsev form 
with $M(k) \propto k^{3/2}$ 
in $k$-space, independent of $\tau$!

\section{Discussion and conclusions}

Fluctuation dynamos, generic to any turbulent plasma, 
are likely to be crucial for rapid generation of magnetic fields
in astrophysical systems.
We have given here an analytical treatment of fluctuation dynamos
at finite correlation times, by modelling the velocity as a flow
which renews itself after every time step $\tau$. 
In particular we present a detailed derivation of the evolution 
equation for the two-point magnetic correlation function in such a flow, 
earlier spelled out briefly in BS14. This generalizes the Kazantsev 
equation which was derived under the assumption that the velocity is 
delta-correlated in time, to the situation where the correlation time is
finite. The correlation time will indeed be finite in any turbulent flow.
Our generalized evolution equation for $M_L(r,t)$ (\Eq{finMleq}), 
reduces to the Kazantsev equation when $\tau \to 0$, and extends it  
to the next order in $\tau$.

The evolution equation for such a finite $\tau$, involves both higher 
(fourth) order velocity correlators and also 
higher order (third and fourth) spatial derivatives of $M_L$, signalling 
that non-local effects are important in this case.
However these higher order derivatives appear only perturbatively, 
multiplied by the small parameter $\bar\tau = \tau\eta_t q^2$. 
This allows us to use the Landau-Lifshitz approach, 
earlier used to treat the effect of the radiation reaction force
in electrodynamics.
In this approach, to the zeroth order in $\bar\tau$, one retains the standard
Kazantsev equation. This is then used to express the third and fourth
derivatives of $M_L$ in terms of the lower order derivatives,
to finally get an evolution equation which at most
involves second derivatives of $M_L$. 

The resulting evolution equation is analyzed both using a
scaling solution and the WKBJ approximation. The scaling solution
is valid in the range of scales, where resistivity can be neglected, 
while the WKBJ treatment also takes into account the effect of a finite
resistivity. From both treatments we see that the effect of a 
finite $\tau$ is to cause a reduction in the dynamo growth rate. 
The asymptotic form of the correlation function on scales
$z_\eta \ll z \ll 1/q$ is very nearly a power law, $M_L \propto z^{-5/2}$ 
independent of $\tau$! This leads to 
the important and intriguing result that the Kazantsev spectrum of 
$M(k) \propto k^{3/2}$, is preserved even at finite-$\tau$. 

Although we derived the effects of a finite-$\tau$ using a particular
renewing velocity field, the resulting evolution equation 
for $M_{ih}$ (\Eq{finalcor}) or $M_L$ (\Eq{finMleq}), can be cast
completely in terms of the general velocity 
correlators, $T_{ij}$ and $T_{ijkl}$. It also matches exactly with 
Kazantsev equation for the $\tau \to 0$ case. 
Moreover, we expect the forms of $T_{ij}$ and $T_{ijkl}$ 
at $r \ll 1/q$, to be universal due to their symmetries and 
divergence free properties.
We would therefore conjecture that our results on the magnetic spectrum
could have a more general validity 
than the context (of a renewing velocity) in which it is derived.
Future work would involve a numerical study of \Eq{finMleq} without
making the small $z$ approximation.
The general methodology developed here also hold the promise
of being systematically extendable to the non-perturbative regime
of $St \sim 1$, at least by a series of numerical
integrations to implement the averaging. The inclusion of shear
and helicity are also the next obvious extensions that need
to be studied, issues which we hope to address in the future. 

\acknowledgments

We thank Dmitry Sokoloff for very helpful correspondence, 
Axel Brandenburg, Nishant Singh and S. Sridhar for several 
useful discussions. PB acknowledges support from CSIR.

\bibliographystyle{jpp}
\bibliography{reftau}

\appendix
\section{Tables for tracking isotropic and homogeneous fourth order tensors}
\label{app1}

\begin{table}
\fontsize{7}{8.4}\selectfont
\setlength{\tabcolsep}{3.5pt}
\caption{The basis tensor components for all fourth order tensors involved in 
\Eq{p1cont}}
\centering
 \begin{tabular}{|c|c|c|c|c|c|c|}
\hline
\hline
Terms & $\left[\rv{ih}\tilde{T}_{mnih}\right]_{,jl}$ & $-\left[\delta_{ij}r_h\tilde{T}_{mnih}\right]_{,l}=$ &
  $-\left[\delta_{il}r_h\tilde{T}_{mnih}\right]_{,j}=$ & $\left(\delta_{ij}\delta_{hl}\right)\tilde{T}_{mnih}$&
Sum/$r^2$ & $M_{jl,mn}$\\ 
$~~$ & $~~$ & $-\left[\delta_{hj}r_i\tilde{T}_{mnih}\right]_{,l}$ &  $-\left[\delta_{hl}r_i\tilde{T}_{mnih}\right]_{,j}$ & $~~$ & $~~$ & $~~$\\ 
\hline
$\rv{jlmn}$ &$r^2\TL^{''}+$ &$-\TL^{'}r-\TL$ &$-\TL^{'}r-\TL$ & $2\TL$ & $\TL^{''}$ & $M_L^{''}$  \\
$~~$ &$4\TL^{'}r+2\TL$ &$~~$ &$~~$ & $~~$ & $~~$ & $~~$  \\
\hline
$\PP{jl}\rv{mn}$ &$\TL^{'}r+2\TN$ &$-\TL+2\TN$ &$-\TL+2\TN$ & $2\TN$ & $\frac{\TL^{'}}{r}-$
& $2M_L^{''} +$ \\
$~~$ &$~~$ &$~~$ &$~~$ & $~~$ & $\frac{(4\TL-12\TN)}{r^2}$
& $\frac{rM_L^{'''}}{2}$ \\
\hline
$\PP{ml}\rv{jn},$ &$(\TL^{'}-\TN^{'})r+$ & $-\TL+2\TN$ &$-\TN^{'}r-\TN$ &
$2\TN$ & $\frac{(\TL^{'}-3\TN^{'})}{r}$ & $\frac{-M_L^{''}}{2}$ \\ 
$\PP{nl}\rv{mj}$ &$(\TL-\TN)$ & $~~$ &$~~$ &
$~~$ & $-\frac{(\TL-3\TN)}{r^2}$ & $~~$ \\ 
\hline
$\PP{nj}\rv{ml},$ & $(\TL^{'}-\TN^{'})r+$& $-\TN^{'}r-\TN$ &$-\TL+2\TN$ & $2\TN$&
$\frac{(\TL^{'}-3\TN^{'})}{r}$ & $\frac{-M_L^{''}}{2}$ \\ 
$\PP{mj}\rv{ln}$ & $(\TL-\TN)$ &$~~$ &$~~$ &
$~~$ & $-\frac{(\TL-3\TN)}{r^2}$ & $~~$ \\ 
\hline
$\PP{mn}\rv{jl}$ & $\TN^{''}r^2+$ & $-\TN^{'}r-\TN$ & $-\TN^{'}r-\TN$ & 
$2\TN$ & $\TN^{''}$ & $\frac{2M_L^{'}}{r}$ \\
$~~$ & $4\TN^{'}r+2\TN$ & $~~$ & $~~$ & 
$~~$ & $~~$ & $~~$ \\
\hline
$\PP{jl}\PP{mn}$ & $\TN^{'}r+2\TN$ & $-\TN$ & $-\TN$ & $2\TLN$ & $\frac{\TN^{'}}{r}+$
& $\frac{3M_L^{'}}{2r}$ \\
$~~$ & $~~$ & $~~$ & $~~$ & $~~$ & $\frac{2(\TLN-\TN)}{r^2}$
& $+\frac{M_L^{''}}{2}$ \\
\hline
$\PP{mj}\PP{ln},$ & $(\TL-\TN)$ & $-\TN$ & $-\TN$ & $2\TLN$ & $\frac{\TL^{'}-5\TN}{r^2}$ &$\frac{-M_L^{'}}{2r}$ \\
$\PP{nj}\PP{lm}$ & $~~$ & $~~$ & $~~$ & $~~$ & $+\frac{(2\TLN)}{r^2}$ &$~~$ \\
\hline
\hline
\label{tab1}
\end{tabular}
\end{table}

\begin{table}
\fontsize{7}{8.4}\selectfont
\setlength{\tabcolsep}{3.5pt}
\caption{The basis tensor components for all fourth order tensors involved in 
\Eq{p2cont}}
\centering
 \begin{tabular}{|c|c|c|c|}
\hline
\hline
Terms & $\left[\rv{h}T_{hnmr}\right]_{,l}$ & $-\delta_{lh}T_{hmnr}$ &
Sum/$r^2$ \\
\hline
$\rv{lmnr}$ &$\TL^{'}r+\TL$ &$-\TL$ &$\frac{\TL^{'}}{r}$  \\
\hline
$\PP{ln}\rv{mr},\PP{lr}\rv{nm},\PP{lm}\rv{rn}$ &$\TL-2\TN$ &$-\TN$ &$\frac{(\TL-3\TN)}{r^2}$\\
\hline
$\PP{mr}\rv{ln},\PP{mn}\rv{lr},\PP{rn}\rv{lm}$ &$\TN^{'}r+\TN$ &$-\TN$ &$\frac{\TN^{'}}{r^2}$\\
\hline
$\PP{mr}\PP{ln},\PP{mn}\PP{lr},\PP{rn}\PP{lm}$ &$\TN$ &$-\TLN$ &$\frac{(-\TLN+\TN)}{r^2}$\\
\hline
\hline
\end{tabular}

\begin{tabular}{|c|c|c|c|c|c|}
\hline
\hline
Terms & $\left[\rv{j}M_{jl}\right]_{,rmn}$ & $-\left(\delta_{jr}M_{jl,mn}\right)$&
$-\left(\delta_{jn}M_{jl,mr}\right)$& $-\left(\delta_{jm}M_{jl,rn}\right)$&
Sum/$r^2$ \\
\hline
$\rv{lmnr}$ &$M_L^{'''}r+3M_L$ &$-M_L^{''}$ &$-M_L^{''}$ &$-M_L^{''}$ &$rM_L^{'''}$  \\
\hline
$\PP{ln}\rv{mr}$ & $M_L^{''}$ & $\frac{M_L^{''}}{2}$ & $-2M_L^{''}-\frac{M_L^{'''}r}{2}$ 
& $\frac{M_L^{''}}{2}$ & $-\frac{M_L^{'''}r}{2}$\\
\hline
$\PP{mr}\rv{ln}$ & $M_L^{''}$ & $\frac{M_L^{''}}{2}$ & $-\frac{-2M_L^{'}}{r}$ 
& $\frac{M_L^{''}}{2}$ & $2M_L^{''}-\frac{2M_L^{'}}{r}$\\
\hline
$\PP{lm}\rv{nr}$ & $M_L^{''}$ & $\frac{M_L^{''}}{2}$ & $\frac{M_L^{''}}{2}$ 
& $-2M_L^{''}-\frac{M_L^{'''}r}{2}$ & $-\frac{M_L^{'''}r}{2}$\\
\hline
$\PP{nr}\rv{lm}$ & $M_L^{''}$ & $\frac{M_L^{''}}{2}$ & $\frac{M_L^{''}}{2}$ 
& $-\frac{-2M_L^{'}}{r}$ & $2M_L^{''}-\frac{2M_L^{'}}{r}$\\
\hline
$\PP{lr}\rv{mn}$ & $M_L^{''}$ & $-2M_L^{''}-\frac{M_L^{'''}r}{2}$ & $\frac{M_L^{''}}{2}$ 
& $\frac{M_L^{''}}{2}$ & $-\frac{M_L^{'''}r}{2}$\\
\hline
$\PP{mn}\rv{lr}$ & $M_L^{''}$ & $-\frac{-2M_L^{'}}{r}$ & $\frac{M_L^{''}}{2}$ 
& $\frac{M_L^{''}}{2}$ & $2M_L^{''}-\frac{2M_L^{'}}{r}$\\
\hline
$\PP{mr}\PP{ln}$ & $\frac{M_L^{'}}{r}$ & $\frac{M_L^{'}}{2r}$ & $-\frac{3M_L^{'}}{2r}-\frac{M_L^{''}}{2}$
&$\frac{M_L^{'}}{2r}$ & $\frac{M_L^{'}}{2r}-\frac{M_L^{''}}{2}$ \\
\hline
$\PP{lm}\PP{rn}$ & $\frac{M_L^{'}}{r}$ & $\frac{M_L^{'}}{2r}$ & $\frac{M_L^{'}}{2r}$ 
&$-\frac{3M_L^{'}}{2r}-\frac{M_L^{''}}{2}$ & $\frac{M_L^{'}}{2r}-\frac{M_L^{''}}{2}$ \\
\hline
$\PP{mn}\PP{lr}$ & $\frac{M_L^{'}}{r}$ & $-\frac{3M_L^{'}}{2r}-\frac{M_L^{''}}{2}$ & $\frac{M_L^{'}}{2r}$ 
&$\frac{M_L^{'}}{2r}$ & $\frac{M_L^{'}}{2r}-\frac{M_L^{''}}{2}$ \\
\hline
\hline
\label{tab2}
\end{tabular}
\end{table}

\begin{table}
\fontsize{7}{8.4}\selectfont
\setlength{\tabcolsep}{3.5pt}
\caption{The basis tensor components for all fourth order tensors involved in 
\Eq{p3cont}. Note that here $\tilde{T}_{mnrs}$ is as in \Eq{expandT}}
\centering
 \begin{tabular}{|c|c|c|c|c|}
\hline
\hline
Terms & $\left[\rv{j}\rv{i}M_{jl}\right]_{,mnrs}$ & $-2\left[\rv{l}M_{(ml}\right]_{,nrs)}$&
$2M_{mn,rs}$ & Sum/$r^2$ \\
\hline
$\rv{mnrs}$ & $M_L^{''''}r^2+8M_L^{'''}r$ & $-8M_L^{'''}r -24M_L^{''}$ & $12M_L^{''}$ & $M_L^{''''}$\\
$~~$ & $+12M_L^{''}$ & $~~$ & $~~$ & $~~$\\
\hline
$\PP{ln}\rv{mr},\PP{lr}\rv{nm},\PP{lm}\rv{rn}$ & $M_L^{'''}r+4M_L^{''}$ & $-8M_L^{''}$ & $M_L^{'''}r+\frac{4M_L^{'}}{r}$ 
& $ \frac{2M_L^{'''}}{r}-\frac{4M_L^{''}}{r^2}$ \\
$\PP{mr}\rv{ln},\PP{mn}\rv{lr},\PP{rn}\rv{lm}$ & $~~$ & $~~$ & $~~$ & $+\frac{4M_L^{'}}{r^3}$ \\
\hline
$\PP{mr}\PP{ln},\PP{mn}\PP{lr},\PP{rn}\PP{lm}$ & $M_L^{''}+\frac{3M_L^{'}}{r}$ & $-\frac{8M_L^{'}}{r}$
& $2M_L^{''}+\frac{2M_L^{'}}{r}$ & $\frac{3M_L^{''}}{r^2}-\frac{3M_L^{'}}{r^3}$ \\
\hline
\hline
\label{tab3}
\end{tabular}

\end{table}

\end{document}